\documentclass[final,5p,times,twocolumn]{elsarticle}

\usepackage{graphicx}
\usepackage{bm}
\usepackage{epsfig}

\usepackage{amssymb}

\newcommand{\Srn}{SrFe$_{2-x}$Ni$_x$As$_2$}
\newcommand{\Src}{SrFe$_{2-x}$Co$_x$As$_2$}
\newcommand{\Ban}{BaFe$_{2-x}$Ni$_x$As$_2$}
\newcommand{\Bac}{BaFe$_{2-x}$Co$_x$As$_2$}
\newcommand{\Srt}{SrFe$_{2-x}$M$_x$As$_2$}
\newcommand{\Sr}{SrFe$_2$As$_2$}

\newcommand{\Niop}{SrFe$_{1.85}$Ni$_{0.15}$As$_2$}

\newcommand{\tc}{$T_c$}

\newcommand{\ie}{{\it i.e.}}

\begin{document}

\begin{frontmatter}



\title{Annealing effects on superconductivity in \Srn}


\author{S.~R.~Saha, N.~P.~Butch, K.~Kirshenbaum, Johnpierre~Paglione}

\address{Center for Nanophysics and Advanced Materials, Department of Physics,
University of Maryland, College Park, MD 20742, USA}

\begin{abstract}
Superconductivity has been explored in single crystals of the
Ni-doped FeAs-compound \Srn\ grown by self-flux solution method. The
antiferromagnetic order associated with the magnetostructural
transition of the parent compound \Sr\ is gradually suppressed with
increasing Ni concentration $x$ and bulk-phase superconductivity
with full diamagnetic screening is induced near the optimal doping
of $x=0.15$ with a maximum transition temperature \tc $\sim$ 9.8~K.
An investigation of high-temperature annealing on as-grown samples
indicate that the heat treatment can enhance \tc\ as much as $\sim
50\%$.

\end{abstract}


\end{frontmatter}


\label{int} The discovery of high-temperature superconductivity in
new iron-based pnictide compounds has attracted much recent
attention~\cite{Saha1}. Suppression of the magnetic/structural phase
transition, either by chemical doping or high pressure, is playing a
key role in stabilizing superconductivity in the ferropnicitides.
Oxygen-free FeAs-based compounds with the ThCr$_2$Si$_2$-type (122)
structure exhibit superconductivity with \tc\ as high as 25 K by
partial substitution of Fe with other transition metal elements,
e.g., \Bac ~\cite{Sefat,Chu,Ni}, \Src ~\cite{Leithe},
\Ban~\cite{Li,Canfield}, \Srt\ (M= Rd, Ir, and Pd) ~\cite{Han}.
Interestingly, in \Bac~\cite{Chu,Ni}, the maximum \tc\ is found at
$x\simeq$0.17, whereas in \Ban, the maximum \tc\ occurs at
approximately $x=$0.10~\cite{Li,Canfield}, suggesting that Ni
substitution may indeed contribute twice as many $d$-electrons to
the system as Co.

We have synthesized and studied single-crystalline \Srn\ and found
that Ni substitution induces bulk superconductivity. Contrary to
expectations framed by prior studies of similar
compounds~\cite{Chu,Ni,Li,Canfield}, we observe a relatively low
maximal \tc\ value of $\sim 10$~K in this series, centered at a Ni
concentration approximately half that of the optimal Co
concentration in \Src~\cite{Leithe}. We have investigated the effect
of high-temperature annealing on as grown samples. Interestingly,
annealing causes an enhancement of \tc\ as much as $\sim 50\%$.

\label{exp}
Single-crystalline samples of \Srn\ were grown using the FeAs
self-flux method~\cite{Saha1}. The FeAs and NiAs binary precursors
were first synthesized by solid-state reaction of (99.999\% pure)
Fe/Ni powder with (99.99\% pure) As  powders. Then FeAs and NiAs
were mixed with elemental (99.95\% pure) Sr in the ratio
$4-2x$:$2x$:1 in an alumina crucible and heated in a quartz tube
sealed in a partial atmospheric pressure of Ar to 1200$^\circ$C.
Crystals were characterized by X-ray diffraction and
wavelength-dispersive X-ray spectroscopy (WDS). Resistivity ($\rho$)
was measured with the standard four-probe ac method in a commercial
PPMS and magnetic susceptibility ($\chi$) was measured in a
commercial SQUID magnetometer.



\begin{figure}[hbtp]
\begin{center}\leavevmode
\includegraphics[width=8cm]{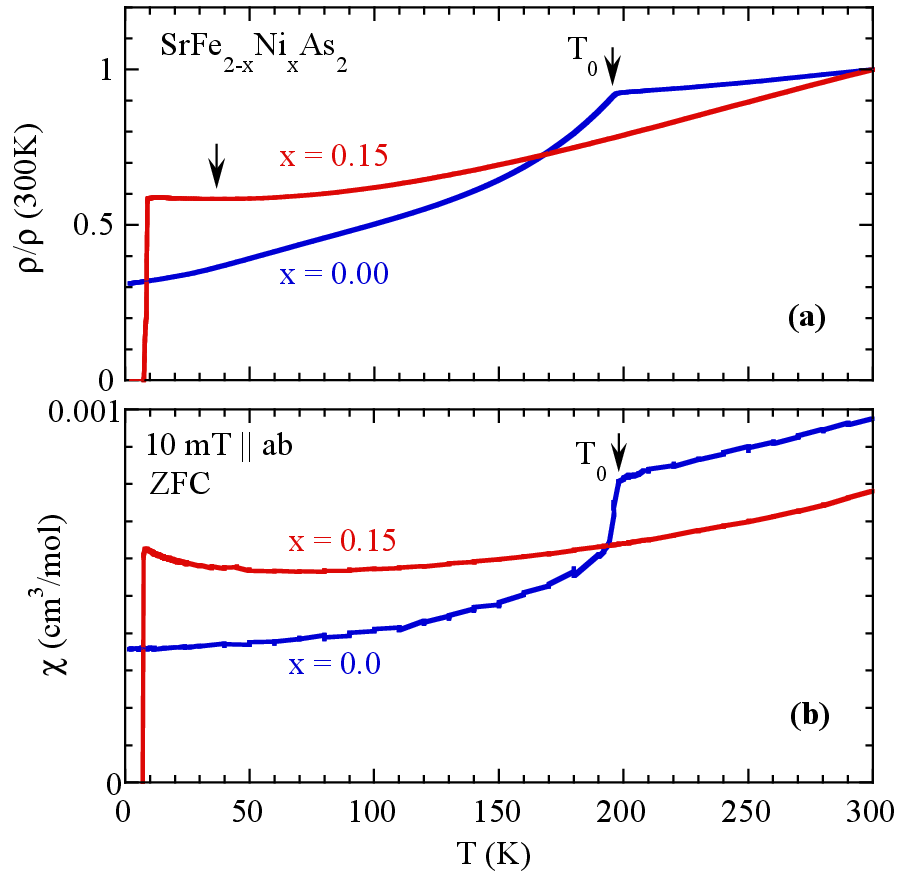}
\caption{(a) Temperature dependence of in-plane electrical
              resistivity in \Sr\ and \Niop\, normalized to 300~K.
              (b) Temperature dependence of magnetic susceptibility $\chi$
              in \Sr\ and \Niop\ for zero-field-cooling (ZFC).
              The arrows indicate the position of $T_0$ (defined in the text).}
\label{fig1}\end{center}\end{figure}


Figure~\ref{fig1}(a) presents the comparison of the in-plane
resistivity $\rho(T)$ between two typical single crystals of \Sr\
and \Niop . As shown, $\rho(T)$ data for \Sr\ decreases with
temperature from 300~K like a metal and then exhibits a sharp kink
at $T_0=198$~K, where a structural phase transition (from tetragonal
to orthorhombic upon cooling) is known to coincide with the onset of
antiferromagnetic (AFM) order~\cite{Yan}. Below $T_0$, $\rho$
continues to decrease without any trace of superconductivity down to
1.8 K. In many undoped \Sr\ samples, strain-induced
superconductivity with $T_c=21$~K has been observed~\cite{Saha2}.
However, here we present $x=0$ data for a sample with all traces of
superconductivity removed by heat treatment. For $x=0.15$, which is
close to optimal doping, the anomaly associated with $T_0$ is
suppressed and transformed into a smooth minimum around 37~K. The
minimum, and hence $T_0$, disappears for $x >$0.15, leading to a
maximum \tc $\sim$ 9.8~K and a dome-like superconducting phase
diagram~\cite{Saha1}. Figure~\ref{fig1}(b) presents the temperature
dependence of the in-plane magnetic susceptibility $\chi$ in \Sr\
and \Niop\ crystals. The overall behavior of $\chi(T)$ for $x=0$
shows a modest temperature dependence interrupted by a sharp drop at
$T_0$. The low-field $\chi(T)$ data at low temperatures presented
here does not show any increase like that in Ref.~\cite{Yan},
indicating both good sample quality (\ie, minimal magnetic impurity
content) and no indication of strain-induced
superconductivity~\cite{Saha2}. For $x=0.15$, the large step-like
feature at $T_0$ disappears and bulk superconductivity is induced
(clearly shown in Fig.~2).

\begin{figure}[hbtp]
\begin{center}\leavevmode
\includegraphics[width=8cm]{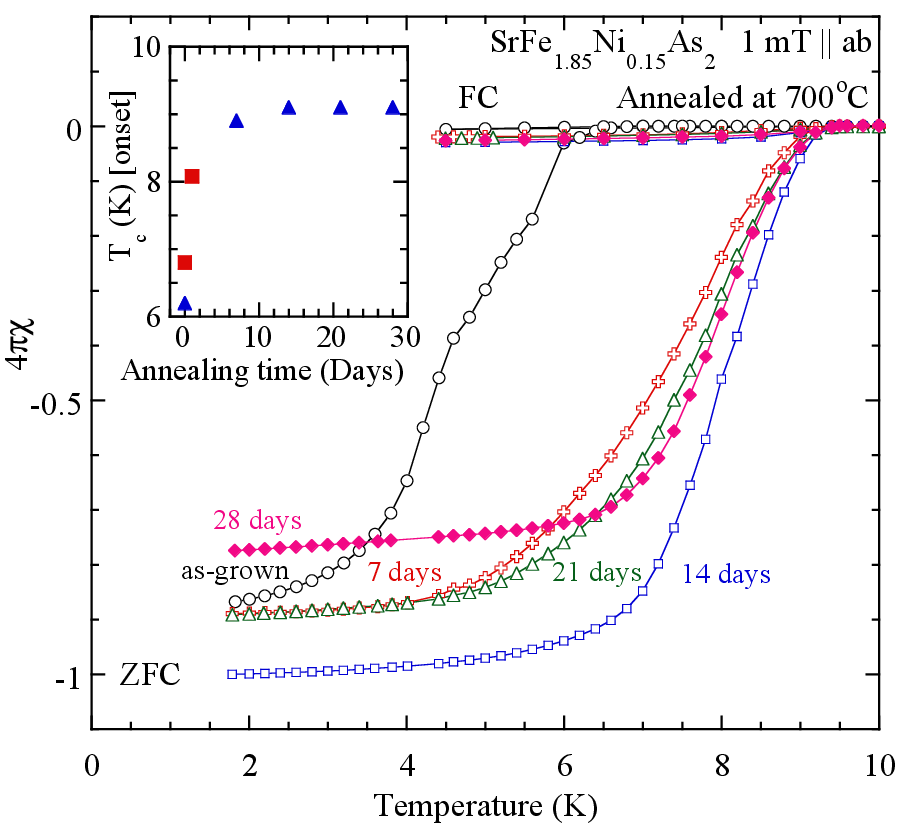}
\caption{Volume magnetic susceptibility in \Niop\ sample measured
before (circles) and after annealing a sample at 700$^\circ$C for 7
days (pluses), 14 days (squares), 21 days (triangles), and 28 days
(diamonds). The lines are guides through the data points. The inset
shows the annealing time dependence of \tc\ for this sample (filled
triangles). The enhancement of \tc\ in a second piece of sample
annealed for 1 day is also plotted (filled squares).}
\label{fig2}\end{center}\end{figure}

We have investigated the effect of high temperature annealing on
single crystals of \Srn\ and found a rather dramatic 10-50\%
enhancement in the value of \tc. This enhancement is reflected in
the full diamagnetic screening and is therefore a bulk phenomenon.
Figure~\ref{fig2} shows the effect of annealing on the
superconducting transition detected in $\chi(T)$ of one \Niop\
annealed at 700$^\circ$C after wrapping with Ta foil and sealing in
a quartz tube under partial atmospheric pressure of Ar. Annealing
for 7 and 14 days enhances the \tc\ (onset) from $\sim$ 6.2 in the
as-grown sample to $\sim$ 8.9~K and $\sim$ 9.2~K, respectively, with
the sharpening of the transition. Annealing for 21 and 28 days does
not enhance the \tc\ further, while it gradually reduces the
superconducting volume fraction, indicating 14 days as the optimal
annealing time. The inset shows the annealing time dependence of \tc
. Enhancement of \tc\ due to  annealing of as-grown \Srn\ (for
several values of $x$) for 1 day at 700$^\circ$C has been found both
in $\rho(T)$ and $\chi(T)$ measurements~\cite{Saha1}. Such an
enhancement of \tc\ could be an indication of improved crystallinity
due to release of residual strain, and/or improved microscopic
chemical homogeneity of Ni content inside the specimens, thereby
optimizing the stability of superconductivity.

A similar annealing effect was reported in $Ln$FeOP (Ln=La, Pr, Nd)
single crystals, where a heat treatment in flowing oxygen was also
found to improve superconducting properties~\cite{baumbach}. It is
further noteworthy to report that some as-grown crystals of \Srn\
for $x<0.16$ (except $x=0.10$) show what looks to be a partial
superconducting transition near 20~K that is completely removed by
heat treatment~\cite{Saha1}. Although it is tempting to posit that
20~K is a possible value for optimal \tc\ in this series of
Ni-substituted compounds, note that aside from the enhancement of
\tc\ as mentioned above, the removal of this feature is the only
change observed in measured quantities imposed by annealing: neither
the normal state resistivity nor magnetic susceptibility show any
change after annealing. Furthermore, susceptibility does not show
any indication of diamagnetic screening in the as-grown samples at
20~K. Because the 20~K kink is removed with heat treatment, and,
moreover, is always found to be positioned near the same
temperature, we believe this feature may be connected to the
strain-induced superconductivity found in undoped \Sr~\cite{Saha2}.
However, note that only a mild 5 minute heat treatment of
300$^\circ$C removes the partial volume superconductivity in \Sr,
while a substantially higher temperature (700$^\circ$C) is required
to remove this feature in \Srn. If the two phenomena are related, it
is possible that internal strain is stabilized by the chemical
inhomogeneity associated with transition metal substitution in \Srn\
thus requiring higher temperatures to remove. More systematic
studies of the effect of annealing on \Srn\ are under way to
investigate the microscopic change in the sample.

%
In summary, single crystals of the Ni-substituted series \Srn\ were
successfully synthesized. The magnetostructural order is suppressed
and bulk superconductivity arises near the optimal doping $x=0.15$
with a \tc\ value reaching as high as $\sim$ 9.8 K. Interestingly,
annealing treatments of as-grown single crystals result in a rather
strong enhancement of the superconducting transition across this
range of $x$, with $\sim 50\%$ increase in \tc\ values for $x=0.15$
for an optimal annealing time of 14 days.

\vskip 0.2cm
%
%
The authors acknowledge P.~Y.~Zavalij and B.~W.~Eichhorn for
experimental assistance, and R.~L.~Greene for useful discussions.
N.P.B. acknowledges support from CNAM.



\end{document}